\documentclass[12pt]{article}

\usepackage{amssymb,amsmath}

\numberwithin{equation}{section}

\include{PhysNote}
\begin{document}

\begin{titlepage}
%\begin{flushright}
%CERN-TH-PH/2012-075
%\end{flushright}
\vskip 1in

\begin{center}

{\Large {\bf Studying Quantum Field Theory}
\footnote{Extended version of an evening talk at the Conference ``Quantum Field Theory, Periods and Polylogarithms'', dedicated to David Broadhurst's 65th birthday, Humboldt University, Berlin, June 2012. Updated, Sects. 2 and 5 made the essential part of an invited talk at the Second Bulgarian National Congress of Physics, Sofia, September 2013. Published in: \textit{Bulg. J. Phys.} {\bf 40} (2013) 93-114.}}

\vspace{4mm}

Ivan Todorov

Institut des Hautes Etudes Scientifiques, 35, route de Chartres
F-91440 Bures-sur-Yvette, France\\ 
e-mail: ivbortodorov@gmail.com

Permanent address:

Institute for Nuclear Research and Nuclear Energy\\
Tsarigradsko Chaussee 72, BG-1784 Sofia, Bulgaria\\
e-mail: todorov@inrne.bas.bg

\end{center}

\begin{abstract}
%\centerline{Abstract}

The paper puts together some loosely connected observations,
old and new, on the concept of a quantum field and on the properties
of Feynman amplitudes. We recall, in particular, the role of 
(exceptional) elementary induced representations of the quantum 
mechanical conformal group $SU(2,2)$ in the study of gauge fields 
and their higher spin generalization. A recent revival of the (Bogolubov-)Epstein-Glaser approach to position space renormalization is reviewed including an application to the calculation of residues of primitively divergent graphs. We end up with an optimistic 
outlook of current developments of analytic methods in perturbative QFT which 
combine the efforts of theoretical physicists, algebraic geometers 
and number theorists. 

%It reflects a taste for unraveling unexpected links between
%physical practices and mathematical structures, as well as
%my attraction to the conformal group and its "local field
%representations" over the years.

\end{abstract}
\end{titlepage}

\tableofcontents
\smallskip
\noindent {\bf Appendix \, Sums of integrals over simplices yielding a
decomposition of Catalan numbers \hfill 18}

\vfill\eject

\section{Introduction}

\begin{flushright}
{\it Before embarking in philosophy one should learn to calculate.} \\
N.N. Bogolubov, advice to young theoretical physicists.
\end{flushright}

In June 1968 a grand style {\it Symposium on Contemporary Physics} was held in
Trieste celebrating the inauguration of the new building of the Abdus 
Salam International Centre for Theoretical Physics. The most memorable
 lectures of the symposium were those pertaining to an evening series entitled
{\it From a Life in Physics}\footnote{A 76-page booklet, so entitled, prefaced
by Abdus Salam, containing all six evening lectures, was published as ``A
special supplement of the IAEA Bulletin, in English only.''}. It included the
recollections of four Nobel Prize winners, Bethe, Dirac, Heisenberg and Wigner,
all in their sixties. People like Heisenberg
and Dirac, the pioneers of quantum theory, were accustomed to lead the way, to
create new trends, not to follow others' changing fashions. It is sad but true
that the most original minds (including even Einstein!) get at some point out of
touch with the current problems in the theory. We were charmed by the stories of
 the heroic past but were not so much interested in the latest contributions of
the founding fathers. Some less creative theorists, whose predilection
is to help understand the intuitive visions of their predecessors
stand a chance to keep longer abreast of what is going on.

I have been studying quantum field theory (QFT) for over 56 years (starting in 1956,
right after graduating the University of Sofia). My first teacher on the subject
became N.N. Bogolubov whose textbook \cite{BS} appeared in 1957. (In the
course of time my views of the field have been also influenced by the
approaches of Arthur Wightman and Rudolf Haag - well before the books \cite{SW, Ha}
became available.) Subsequently, I was mostly learning from my students and younger
collaborators. With no attempt to exhaust their list let me single out Anatolii Oksak
with whom we studied infinite component fields since the mid 1960's \cite{OT, OT69,
OT70} and who was instrumental in completing the book \cite{BLOT}; Gerhard Mack and
Detlev Buchholz the collaboration with whom \cite{MT, DMPPT} and \cite{BMT} has proved beneficial to our group in Sofia. An outgrow of our joint work with Mack was the study of exceptional elementary representations of the conformal group by Vladimir Dobrev, Valya Petkova and Galen Sotkov, surveyed in Sect. 2.2. A renewal of interest in higher dimensional conformal QFT was started at the turn of the century with my youngest collaborator, Nikolay Nikolov \cite{NT, NT05} and continued with my (once student) long term partner Yassen Stanev (now in Rome)\cite{NST03}, Karl-Henning Rehren \cite{NRT08}, and Bojko Bakalov \cite{BNRT}. Dirk Kreimer taught me - through his work with David Broadhurst at ESI, his spectacular collaboration with Alain Connes, \cite{Kr, CK}, and his later work with Spencer Bloch, \cite{BEK, BK} - both at IHES -  that renormalization is a 
 beautiful and lively subject. I keep learning (and appreciating) it while pursuing
our (still continuing) joint project \cite{NST, NST12} with a veteran in the field, 
Raymond Stora, and with Nikolov.

Following the advice of my first teacher in QFT I'll refrain from philosophizing
and keep my comments down to earth - and close to the calculations.

\bigskip

\section{The role of the conformal group in QFT} 
%Fields as sections of induced $SU(2,2)$ bundles}
\setcounter{equation}{0}
\subsection{Fields versus particles}
There has been a tension between the field and the particle picture in QFT since
 its inception. Particles correspond to unitary positive energy {\it irreducible
representations} (IRs) of the Poincar\'e group. According to the classification
 of Wigner \cite{W39} they are labeled by the mass $m\geq 0$ and the spin (or 
helicity for $m=0$) $s$. They are constructed as representations induced by the
 invariance subgroup of a {\it particle momentum}. Local fields, on the other 
hand, are characterized by their dimension $d$ and by an IR $(j_1, j_2)$ (of 
dimension $(2j_1+1)(2j_2+1)$) of the quantum mechanical Lorentz group 
$SL(2, {\mathbb C})$. Assuming that in QFT we privilege fields, \cite{Ho}, one 
can ask whether there is a representation theoretic relation between the Lorentz
 weight and the scale dimension, and between them and the associated particle 
characteristics. A unexpected answer is suggested by studying the conformal 
symmetry of the limit theory of massless fields. It is provided by the {\it 
elementary induced representations} (EIRs) of the {\it quantum mechanical  
conformal group} ${\mathcal C} = SU(2, 2)$, a four-fold cover of the simple 
"geometric conformal group" $SO_0(4, 2)/\{\pm 1\}$ that appears as the group of
local causal automorphisms of space time (cf. \cite{FS, S}). There are two ways
to describe the EIRs of ${\mathcal C}$, \cite{M, T}. One is to view them as induced 
by irreducible representations (IRs) of the stability subgroup of a point, say $x=0$; 
this is an 11-parameter {\it parabolic subgroup} ${\mathcal P} \subset {\mathcal C}$ 
which can be written as a semidirect product of three factors:
\begin{equation}
\label{MAN}
{\mathcal P} = MAN(= N\rtimes (M\times A))\ , \quad M = SL(2, {\mathbb C})\ , \quad A={\mathbb R}_+\ ,
\end{equation}
$A$ is the 1-parameter subgroup of dilations $x \rightarrow \rho x, \, \rho>0$
commuting with Lorentz transformations and N is the 4-parameter (abelian) subgroup of {\it special conformal transformations}\footnote{As $n_c$ acts on Minkowski space ${\mathbb M}={\mathbb R^{3,1}}$ with singularities one should view (classical) conformal fields as sections of a ${\mathcal C}$ vector bundle on the conformal compactification of ${\mathbb M}$.} $n_c: x \rightarrow (x+x^2 c)(1 + 2xc + x^2 c^2)^{-1}$. The invariant subgroup
$N\subset {\mathcal P}$ (with nilpotent Lie algebra) has only trivial finite dimensional IR. Thus the inducing representations of ${\mathcal P}$ are labeled precisely by the triple $[d; j_1,j_2]$
giving the IRs of ${\mathbb R}_+\times SL(2, {\mathbb C})$ that characterize local fields. The second approach, appropriate to {\it positive energy (highest weight)} representations, uses induction from the IRs of the maximal compact subgroup $K = S(U(2)\times U(2))\in {\mathcal C}$. These are again labeled by triples of the above type with $(j_1, j_2)$ now standing for an IR
of the semisimple part $SU(2)\times SU(2)$ of $K$ while for the EIRs that admit a unitary positive energy subrepresentation $d$ appears as the minimal eigenvalue of the generator $H$ of the centre
$U(1)$ of $K$, interpreted as the {\it conformal Hamiltonian}. {\it Globally conformal invariant} (GCI) fields \cite{NT} transform under proper (rather than projective) EIRs of ${\mathcal C}$ for which the {\it twist} $d-j_1-j_2$ is an integer. Following \cite{T} we shall call such triples $[d; j_1, j_2]$ {\it integer points} in the space of EIRs:
\begin{equation}
\label{EIR}
{\mbox EIR}: \chi = [d; j_1, j_2]\ , \quad {\mbox GCI}\Rightarrow d-j_1-j_2 \in {\mathbb Z}\ .
\end{equation}
Such integer EIRs have the good sense to admit both lowest and highest weight subrepresentations,
thus giving room to (free) fields that admit a decomposition into a positive and a
negative frequency part. Adding the requirement of {\it Hilbert space (or
Wightman) positivity}, i.e. demanding that the space generated by the action of a GCI (smeared quantum) field on the vacuum admits a (non-trivial) subspace realizing a {\it unitary} positive energy irreducible representation of ${\mathcal C}$ we will deduce that the twist should be positive: $d-j_1-j_2=1, 2, ...$. (Twist one only appears for free {\it chiral} fields - i.e., for $j_1 j_2 = 0$.) Remarkably, these are precisely the constraints that single out relativistic fields appearing in a local QFT (when no conformal or scale invariance is assumed, $d$ is to be identified with the naive "canonical" dimension of the field in units of inverse length).

{\footnotesize A digression: massless fields are important for QFT in more than one way. They naturally split into {\it chiral} (irreducible under $SL(2,{\mathbb C})$) components. The recognition that left and right chiral fields interact differently was basic to the formulation of the standard model of particle physics. As noted in (Appendix G to) \cite{MZ12} the understanding of spin one massless fields led to the notion of gauge symmetry and to the theory of Maxwell and Yang-Mills fields; a massless interacting spin two field gives rise to the (invariant under diffeomorphisms) Einstein gravity; higher spin fields (see \cite{FV, V12} and references in the latter paper) may provide a new point of departure for string theory. One can trace a parallel between starting with the zero-mass limit in the fundamental laws of QFT and Galileo's neglecting friction and recognizing the role of inertial frames which paved the way to creating classical mechanics. From this point of view we are still in the formative period of QFT. The hope that masses may emerge without being put in by hand is supported by the idea of dimensional transmutation \cite{F} and by lattice results in "QCD lite" -see \cite{W12}.}

Lorentz invariant (massless) field equations are only conformal invariant for special integer EIRs, namely those corresponding to twist one. This condition is indeed obeyed by the Dirac (or Weyl) and the Maxwell equations. Conserved currents appear just as twist two integer EIRs as in the familiar cases of the electromagnetic current and the stress-energy tensor. They are also intertwined with gauge fields (and their generalizations), as we proceed to demonstrate. 
\smallskip
\subsection{Gauge fields as exceptional EIRs}
Gauge fields and conserved currents turn up as special {\it relatives} (having the same eigenvalues of the Casimir operators) to the finite dimensional IRs\footnote{Here we outline the results of \cite{PS, PST} - an outgrow of \cite{DMPPT, DP, KG, T81}.}.

Every finite dimensional IR appears as a subrepresentation of an
EIR $\chi_{-\nu}$ for which $d+j_1+j_2 =:1-\nu$ is a non-positive
integer (so that $\nu = 1, 2, ...$). It is convenient to express
the labels $d, j_1, j_2$, in terms of $\nu$,  the total spin $\ell
= j_1+j_2$ and another positive integer $n = 2j_2 + 1\ (=1, ...,
2\ell+1)$. ($2j_2$ being the number of symmetrized dotted $SL(2, {\mathbb C})$-indices.) There are six partially equivalent EIRs, which we label by the value of $d + \ell - 1$:  
$\chi_{-\nu} = [1-\ell-\nu; \ell - \frac{n-1}{2}, \frac{n-1}{2}], \chi_0 = [1-\ell;\ell + \frac{\nu-n+1}{2},\frac{\nu+n-1}{2}],
\chi_n=[1-\ell + n; \frac{\nu+1}{2}, \frac{\nu-1}{2}]$ and their duals ${\tilde \chi}_{-\nu}
(= [3+\ell+\nu, \frac{n-1}{2}, \ell-\frac{n-1}{2}]), {\tilde \chi}_0, {\tilde \chi}_n$ which can be combined in two exact sequences (where we identify the EIRs $\chi$ with the corresponding space-time bundles): 
%starting with $\chi_{-\nu}$ that are relatives to it, as displayed below:
%{\scriptsize
\begin{eqnarray}
&&(0 \rightarrow) \chi_{-\nu} %= [1-\ell-\nu; \ell - \frac{n-1}{2}, \frac{n-1}{2}]
\rightarrow \chi_0 %= [1-\ell;\ell + \frac{\nu-n+1}{2}, \frac{\nu+n-1}{2}]
%%\nonumber \\&&
\rightarrow \chi_n (\rightarrow 0) \qquad \nonumber \\
%=[1-\ell + n; \frac{\nu+1}{2}, \frac{\nu-1}{2}] \qquad \nonumber \\
&&\label{Exc}\\
&&(0 \leftarrow ){\tilde \chi}_{-\nu}%=[3+\ell +\nu; \frac{n-1}{2}, \ell -\frac{n-1}{2}] 
\leftarrow {\tilde \chi}_0 \leftarrow {\tilde \chi}_n (\leftarrow 0).
%= [\ell + 3; \frac{\nu+n-1}{2}, \ell + \frac{\nu-n+1}{2}] %%\nonumber \\&&
%\leftarrow {\tilde \chi}_n=[\ell +3 -n; \frac{\nu-1}{2},\frac{\nu+1}{2}]\ .
\nonumber
\end{eqnarray}
The arrows indicate differential intertwining operators of order
$\nu$ and $n$, respectively (and the image of each map is the kernel of the 
subsequent). The dual EIRs ${\tilde \chi}$  (with ${\tilde d}=4-d$ and $j_1$ 
and $j_2$ interchanged) 
of the second line are partially equivalent to the $\chi$'s above them. There 
are also differential intertwining maps of order $2\ell+2-n$ (not indicated
above) $\chi_0 \rightarrow {\tilde \chi}_n$ and $\chi_n\rightarrow
{\tilde \chi}_0$. For $(\nu, \ell, n) = (1, 0, 1)$ we obtain the
equations related to a gauge field. The arrow on the top row pointing to 
$\chi_0$ is the gradient, mapping a dimensionless scalar function $s$ to a
pure gauge field $\partial_\mu s$. The two arrows originating in
$\chi_0$ map the 4-vector $A_\mu$ (of dimension 1) onto the
(anti)selfdual projections of its curl. The two arrows ending in
${\tilde \chi}_0$ and the one originating there are divergences;
the vanishing of the map ${\tilde \chi}_0 \rightarrow {\tilde
\chi}_{-\nu}$ on the image of the two preceding maps expresses the
conservation of the Maxwell current. One may speculate about the
relation of other exceptional EIRs to the gauge theory of higher
spin fields. In particular, the sextuplet labeled by $(\nu, \ell,
n) = (1, 1, 2)$ appears to be related to a conformal gauge theory
of gravity. More generally, we can single out the series of symmetric tensor representations
corresponding to $\nu = 1, \, n= \ell + 1, \, \ell = 0, 1, ...$. They have vanishing higher order Casimir invariants\footnote{In general, $C_2 = (\ell+1-n)^2 + \ell(\ell + 2) + (\ell + \nu +3)(\ell + \nu - 1), \, C_3=(d-2)(j_1-j_2)(j_1+j_2+1)= -(\ell+1)(\ell+1-n)(\ell+1+\nu),\, C_4 = n^2 (2\ell+2-n)^2 +1 - 2[(\ell+1)^2 + (\ell+1-n)^2][(\ell+\nu+1)^2 + 1] +(\ell+\nu+1)^2 [(\ell+\nu+1)^2 - 2] (=0$ for $n=\ell+1, \, \nu=1)$.}:
\begin{equation}
\label{Cas}
C_2(1, \ell, \ell+1) = 2 \ell (\ell + 3), \, C_3 = 0 = C_4.
\end{equation}

{\footnotesize The above construction extends to arbitrary space-time dimension
 $D$ [KG]. We shall summarize the case of even $D=2h$ in which there are $D+2$ 
relatives of any finite dimensional IR (see also Appendix B.4 of [TMP]; for odd
 $D$ the number of relatives is $D+1$). The EIRs of the 4-fold cover $Spin(D,2)$
 of the D-dimensional conformal group are labeled by $h+1$ numbers: $\chi = 
(d; l_1, ..., l_h)$ where $l_1 - l_2, ..., l_{h-1}-|l_h|$ are non-negative 
integers. The dual representation ${\tilde \chi}=(D-d; l_1, ..., l_{h-1}, -l_h)$
 contains a finite dimensional subrepresentation if $\nu:=d-D-l_1+1$ is a 
positive integer. Setting $\chi_h = (l_h+h, l_1+\nu, l_1 +1, ..., l_{h-1}+1), 
..., \chi_{2h-1} = (2h-1 + l_1, l_1+\nu, ..., l_h), \chi_{2h} = (2h+l_1+\nu -1, 
l_1, ..., l_h)$, we can write the counterpart of the second row of (\ref{Exc}) 
as $\chi_h \rightarrow \chi_{h+1}\rightarrow ... \rightarrow \chi_D$. In 
particular, for $\nu=1$ the kernel of the last arrow corresponds to the twist 
$D-2$ conseved (tensor) currents. The case $h = 2$ is recovered for 
$l_1=\ell \, (=j_1+j_2),\, l_2=j_1-j_2$.} 
\smallskip
\subsection{Higher spin conserved currents}
 
So far we only came across a class of representations of the
conformal group  appearing in the beginnings of field theory.
Let me allude to a couple of more advanced - and more recent - applications.

The wealth of 2-dimensional conformal field theories (CFT)  uncovered by 
Belavin, Polyakov and Zamolodchikov in 1984 (preceded by the lonely Thirring 
model of 1958), stimulated attempts to extend at least some of the results 
and techniques to higher dimensions. Nikolov and I tried to
generalize the notion of a (chiral) vertex algebra by exploiting
the concept of global conformal invariance \cite{NT, N, NT05}.
This worked smoothly in even space time dimensions and led to CFT
models with Huygens locality, rational correlation functions and
an infinite number of conservation laws. Recent work of Maldacena
and Zhiboedov \cite{MZ}, motivated by the $AdS_4/CFT_3$
correspondence, considered a 3D CFT with a higher spin symmetry.
They reproduce in $3D$ (with no Huygens locality) an argument and part of the 
results of \cite{BNRT} stating that (for $D=3$) any theory with an infinite number 
of conserved currents is generated by normal products of free fields. This can be
viewed as an extension of the Coleman-Mandula theorem [CM67] to the case of 
a CFT (a theory with no mass gap and, typically, no scattering matrix). The role of 
the S-matrix as a basic observable is replaced by the correlation functions of the 
stress-energy tensor - as suggested long ago by Mack \cite{M77} (and elaborated recently by a number of authors - see \cite{Z13, BHKSZ} and references therein). All this made me 
think that the proper generalization of a ($2D$) chiral algebra,
valid for both even and odd dimensions, is precisely a CFT with an
infinite set of conserved currents \cite{T12}. Conserved (symmetric) tensors in
 $D$ space-time dimensions appear for twist $D-2$ (twice that of a free 
massless field). The attention towards theories with higher conservation laws
stimulated a renewed interest in the study of conformal 3-point
functions of conserved (tensor) currents (see an important early
contribution by Stanev \cite{S88}, demonstrating that the general 3-point function of the stress-energy tensor is a linear combination of three free field structures, and the outcome of later work in \cite{GPY, MZ, St, Z}). It led, in particular, to the following
generalization of the Weinberg-Witten theorem \cite{WW80}: {\it
Let $\phi_s(x)$ be a massless (free) field that transforms under
the representation\footnote{The corresponding UPEIRs of $U(2,2)$,
describing massless particles of any helicity were proven to
remain irreducible when restricted to its Poincar\'e subgroup
\cite{MT69}. Note that these IRs do {\it not} belong to the exceptional 
series described in Sect. 2.2.} $[s+1; s, 0] (s=0, 1/2, 1, ...)$ of ${\mathcal C})$.
The necessary and sufficient condition for the existence of a
conserved rank $r$ (and twist two) tensor current $J_r(x)$, such
that the conformal invriant 3-point function
$<\phi_s(x_1)\phi_s(x_2)^*J_r(x_3)>$ does not vanish, is $r\geq
2s$} \cite{St, T12} ($\phi_s^*$ transforming under $[s+1; 0, s]$).

The Thirring model was a source of inspiration for introducing (by Wilson and 
others) the notion of {\it anomalous dimension} signaling the use of projective
 representations of the conformal group (or, equivalently, representations of 
the infinite-sheet universal cover of ${\mathcal C}$). The now fashionable 
${\mathcal N}=4$ superconformal Yang-Mills model also involves fields of 
anomalous dimensions and may be viewed as a far going $4D$ generalization of 
the Thirring model. Mack and I constructed, back in the 1970's \cite{MT}, a 
skeleton QFT with conformal invariant dressed vertex functions and propagators 
(involving anomalous dimensions) that is free of divergences (work later 
surveyed in \cite{TMP, T07}). A closely related work led us, together with my students, to the study of {\it conformal partial wave expansions} \cite{DMPPT}. A fresh retrospective view of these developments  and their relation to the "dual resonance models", provided by Mack \cite{M09}, is currently taken up by a young Portuguese team \cite{CGP}.
%(Some six years ago I had an occasion to review this work in more detail, \cite{T07}.)
\bigskip

\section{Renormalization in configuration space}
\setcounter{equation}{0}
\subsection{Wightman versus time-ordered Green functions}

Quantum field theory encountered serious problems - the appearance of divergent
 integrals in perturbation theory - from its inception. The founding fathers were
even ready to abandon it, anticipating yet another revolution in the theory. A
new generation of pragmatically minded physicists on the wake of the war used
instead judiciously the renormalization ideas of H. Kramers and others of
 the 1930's, transforming them into concrete recipes for getting finite numbers
 out of the mess. Moreover, these numbers gave amazingly good fits to high
precision experimental results. The problem arose for mathematically inclined
theorists to disentangle sense from nonsense and provide a coherent picture
of QFT. If the practical S-matrix calculations (including the mass, charge and
wave function renormalization) were performed in momentum space, nearly all
attempts to give a precise formulation of the basic principles of QFT are
using the position space picture. The oscillation between the position and
momentum space approaches is a trademark of perturbative renormalization (an
early manifestation of the now fashionable concept of duality - realized by the
 good old Fourier transform). While the notion of a {\it causal} S-operator was
introduced by Ernst Stueckelberg and his collaborators (D. Rivier, T.A. Green,
A. Petermann) back in 1950-53 and taken up by Bogolubov and his school (as 
reviewed in \cite{BS}), A.S. Wightman proposed his version of axiomatic QFT 
(see \cite{SW, BLOT} and references therein). It provides an instructive example
 of the interplay between a nice mathematical outlook and the stuff needed by 
physicists.

Wightman distributions are boundary values of analytic functions in a tube
domain\footnote{I only learned recently, reading a "golden oldie" \cite{W}, that 
the idea to consider Green functions in the upper half-plane of the eigenvalue 
parameter "not as trivial at that time as it has now become" goes back to a 1910 
paper by Hermann Weyl.} and can be multiplied. As a result, correlation functions 
of normal products of derivatives of free fields are well defined distributions. 
In a dilation invariant (massless) QFT they are dilation covariant (homogeneous) 
boundary values of elementary functions. They form a graded (by the degree of 
homogeneity) algebra generated by 2-point functions of the type
\begin{eqnarray}
\label{2pt}
&&W_{12} = \frac{P(x_{12})}{\rho_{12}^m}\ ,\quad x_{12} = x_1 - x_2 \ , \quad  \rho_{12} = (x_{12}^2+i0x_{12}^0)^{1/2}, \nonumber \\
&&x^2 = {\bf x}^2 - (x^0)^2 \ , \qquad {\bf x}^2 = \sum_{i=1}^{D-1} (x^i)^2
\end{eqnarray}
(modulo obvious relations); here $P=P_{12}$ is a homogeneous polynomial and $m = m_{12}$ is a positive integer (that is even for even space-time dimension D). A nice object - and no
divergences in sight - should not everybody be happy? But, in order to set up a
perturbation theory allowing to compute physical quantities - scattering amplitudes,
form factors, magnetic moments - one needs {\it time-ordered Green functions} (that are invertible - unlike the Wightman functions - under the convolution product). Here resurges the  problem of ultraviolet divergences albeit in a tamed and more respectable form: as a problem of extension of distributions.\footnote{We are not going to touch upon other complications involving large distance behaviour: existence of adiabatic asymptotic limits, control over mass shell singularities - for whose study the momentum space picture still appears to be more appropriate.} A simple
 example of what is involved is provided by (the powers of) the Wightman and the
 time ordered 2-point functions of a free massless scalar field in $4D$:
\begin{eqnarray}
\label{TW}
&&G(x_{12}) = \, <0|T(\varphi(x_1)\varphi(x_2))|0>\, = \left\{\begin{array}{ll}
w(x_{12})&{\rm for}\  x_{12}\notin {\bar V}_-  \\
w(x_{21})&{\rm for}\  x_{21}\notin {\bar V}_-    \,  \,  \,
\end{array}\right.                            \nonumber \\
&&= \frac{1}{4\pi^2(x_{12}^2 + i0)} \qquad({\bar V}_\pm = \{x; \pm x^0 \geq |{\bf x}|\}) \ ,   \nonumber \\
&&w(x_{12}) :=\, <0|\varphi(x_1)\varphi(x_2)|0> \, = \frac{1}{4\pi^2\rho_{12}^2}\ .  \,  \,
\end{eqnarray}
If $n>1$ then the powers of the propagator $G^n(x)$ are only unambiguously
defined for $x\neq 0$.
Their extension to distributions on the entire Minkowski space
${\mathbb M}$ involves a renormalization ambiguity - a distribution with
support at the origin. Moreover, if we demand that the renormalization map is
linear (in particular, that the extension of zero is zero) then it cannot
commute with all derivatives - it has to involve {\it anomalies}. Indeed, while
 $w(x)$ (and $G(x)$ away of the origin) satisfy the d'Alembert equation,
$\square\, w(x)=0$, the time ordered Green's function obeys the inhomogeneous
equation $\square\, G(x)=i\delta(x)$ (where $\delta(x)$ is the 4-dimensional Dirac
$\delta$-function). Thus, had we assumed that renormalization commutes with
derivatives  we would have to violate linearity allowing that the
renormalization of zero may be $i\delta(x)$.

Dualities are chiefly used to reduce a large parameter asymptotics to a small
parameter one. From this point of view it is natural to overturn the habit of
just looking at momentum space, and begin studying ultraviolet - small distance
 - singularities in position space. That was a natural starting point for the
work of Bogolubov - a mathematician set to master QFT - and was implemented
systematically by Epstein and Glaser \cite{EG} (for later developments and
further references - see \cite{BBK}). Some 30 years later when developing
perturbative QFT and writing down operator product expansions on a curved
background became the order of the day, it was realized that it is the x-space
approach which offered a way to their implementation - see \cite{BF, DF, HW}.
There should be, therefore, no surprise that this approach attracts more
attention now than forty years ago when it was conceived. Our interest, 
triggered by Raymond Stora, a reputed master of the field, started with the 
observation that H\"ormander's treatment of the extension of homogeneous 
distributions (Sect. 3.2 of \cite{H}), when generalized to associate 
homogeneous ones, becomes tailor-made for studying the ultraviolet 
renormalization problem; that is particularly transparent in a massless QFT. For
 us, \cite{NST, NST12}, a renormalization is a map (satisfying certain 
conditions) from a space of integrable functions defined on a dense open subset
 of ${\mathbb R}^N$ (outside the singularities) to distributions on ${\mathbb 
R}^N$. This way of reformulating the renormalization problem has given enough 
ground for an invitation \cite{FHS} addressed to sceptical theorists to ``stop 
worrying and love QFT''. The rest of my talk provides some highlights of our 
work.

\smallskip
\subsection{Causal factorization}
Let me start with the basic requirement of {\it causal factorization} which
allows to define the ultraviolet renormalization of an arbitrary graph by
induction in the number of vertices. For the sake of simplicity I shall only
spell out its euclidean version \cite{NST12}.

Assume that all contributions of diagrams with less than $n$ points are renormalized. If then $\Gamma$ is an arbitrary connected $n$-point graph its renormalized contribution should satisfy
the following inductive {\it factorization requirement}. Let the index set $I(n) = \{1,\ldots ,n\}$ labeling the vertices of $\Gamma$ be split into any two non-empty non-intersecting subsets
$$
I(n) = I_1 \dot\cup \, I_2 \quad (I_1 \ne \emptyset \, , \ I_2 \ne \emptyset \, , \ I_1 \cap I_2 = \emptyset) \, .
$$
Let ${\mathcal U}_{I_1,I_2}$ be the open subset of $\mathbb{R}^{Dn} \equiv (\mathbb{R}^D)^{\times n}$ such that $(x_1,\ldots,x_n) \notin {\mathcal U}_{I_1,I_2}$  whenever there is a pair $(i,j)$ such that $i \in I_1$, $j \in I_2$. Let further $G_1^R$ and $G_2^R$ be the contributions of the subgraphs of $\Gamma$ with vertices in $I_1$ and $I_2$, respectively. For each such splitting our distribution $G_{\Gamma}^R$, defined on all partial diagonals, exhibits the {\it euclidean factorization property}:
\begin{equation}
\label{CF}
G_{\Gamma}^R = G_1^R \left(\prod_{i \in I_1 \atop j \in I_2} G_{ij} \right) G_2^R \quad \mbox{on} \quad {\mathcal U}_{I_1,I_2} \, ,
\end{equation}
where $G_{ij}$ are factors (of type (\ref{2pt}) with $\rho_{ij}$ replaced by the
 euclidean distance) in the rational function $G_{\Gamma}$ and are understood as
 {\it multipliers} on ${\mathcal U}_{I_1,I_2}$. This property is inspired by the
 Minkowski space causal factorization of Epstein-Glaser \cite{EG} in which one 
replaces the product of $G_{ij}$ by a product of (equally oriented) Wightman 
functions $W_{ij}$ (\ref{2pt}) and applies the wave-front criterion (\cite{H} 
Chapter VIII) - see the second reference \cite{NST}.

\bigskip

\section{Residues. A finite renormalization}
\setcounter{equation}{0}

\subsection{Renormalization of primitively divergent graphs}

From now on we shall restrict our attention to the case of a massless theory in which Feynman propagators are elementary homogeneous functions like (\ref{TW}). We say that a homogeneous "bare" Feynman amplitude $G(\vec x)$ (corresponding to a connected graph) is {\it (superficially) divergent} if the degree of homogeneity $-\kappa$ of the density $G(\vec x)Vol$ is non-positive. Realizing the inductive procedure suggested by the factorization requirement of the preceding section one starts with
the renormalization of {\it primitively divergent graphs} - i.e. divergent graphs without subdivergences. The following proposition (see Theorem 2.3 of \cite{NST12}) may serve as a definition of both a distribution valued {\it residue} Res and a {\it primary renormalization map} ${\mathcal P}_N^r: {\mathcal S}' ({\mathbb R}^N \backslash \{ 0 \}) \to {\mathcal S}' ({\mathbb R}^N)$.

\smallskip

\noindent{\bf Proposition.}
{\it If $G^0 (\vec x)$ is a homogeneous distribution of degree $-d$ on ${\mathbb R}^N \backslash \{ 0 \}$ $(d=N+\kappa \geq N)$ and $r=r(\vec x)$ is a semi-norm on ${\mathbb R}^N$, then
\begin{equation}
\label{PrG}
r^{\epsilon} \, G^0 (\vec x) - \frac{1}{\epsilon} \, ({\rm Res} \, G)(\vec x) = G^r(\vec x) + O(\epsilon) \quad (G^r = {\mathcal P}_N^r \, G^0) \, ;
\end{equation}
here $G^r$ is the extended - renormalized distribution, ${\rm Res} \, G$ is a distribution with support at the origin whose calculation is reduced to the case $d=N$ of a logarithmically 
divergent graph by using the identity
\begin{equation}
\label{Res}
{\rm Res} \, G = \frac{(-1)^{\kappa}}{\kappa!} \, \partial_{i_1} \ldots \partial_{i_{\kappa}} ({\rm Res} (x^{i_1} \ldots x^{i_{\kappa}} \, G))(\vec x)
\end{equation}
where summation is understood over all repeated indexes $i_1 , \ldots , i_{\kappa}$ from $1$ to $N$. If $G^0 (\vec x)$ is homogeneous of degree $-N$ then
\begin{equation}
\label{lnRes}
{\rm Res} \, G(\vec x) = ({\rm res} \, G^0) \, \delta (\vec x) \qquad (\mbox{for} \ (\vec x \vec \partial + N) \, G^0 (\vec x) = 0)
\end{equation}
where
\begin{equation}
\label{res}
{\rm res} \, G^0 = \int_{r=1} G^0 (\vec x) \sum_{j=1}^N (-1)^{j-1} \, x^j \, dx^1 \wedge \ldots d \widehat x^j \ldots \wedge dx^N
\end{equation}
is independent of the value of $r$ (here taken as one) since the form under the
 integral sign is homogeneous and closed. (A hat, $\widehat{ \ }\,,$ over an 
argument means, as usual, that this argument is omitted.)}

Let me illustrate the computation of a residue with a known example, "the wheel with n spokes" (see \cite{B, BrK} and earlier work of D.J. Broadhurst cited there).  
It is a primitively divergent $4D$ $n$-loop Feynman amplitude. Choosing the origin of the coordinate system at the center of the wheel and labeling the remaining 
vertices by $0, 1, ..., L(=n-1)$ (as in [B]) we can write:
\begin{equation}
\label{Gnloop}
G(x_0, x_1, ..., x_L) = (\prod_{i=0}^L x_{i}^2 x_{i i+1}^2)^{-1}\ ,\quad x_{L+1} \equiv x_0,
\end{equation}
which we shall parametrize by the spherical coordinates of the $n$ independent 4-vectors $x_{i}$:
\begin{equation}
\label{x0i}
x_{i} = r_i \, \omega_i \, , \quad r_i \geq 0 \, , \quad \omega_i^2 = 1 \, , \quad i=0, 1, 2, ..., L\,.
\end{equation}
(For $n=L+1=3$ we recover the complete 4-point graph of the $\varphi^4$ theory.)

Setting
\begin{equation}
\label{Gnr}
G_n^\varepsilon = \left( \frac{r_0}{\ell} \right)^\varepsilon G_n\ , 
%\quad  r_0 \geq max(r_1, ..., r_L),
\end{equation}
we shall compute its residue by first integrating the corresponding analytically regularized density $G_n^{\varepsilon} \mathrm{Vol}$ over the angles 
$\omega_i$, using the old trick \cite{CKT, B, KTV}. of expanding each propagator $\frac{1}{x_{ij}^2}$ with in Gegenbauer polynomials (see Appendix). Taking the homogeneity of the integrand (and the convergence of the $4L$ dimensional integral) into account we find 
\begin{equation}
\label{resBro}
  \, {\rm res} \, G_n = \vert {\mathbb S}^3 \vert r_0^4 \int d^4x_1 \ldots \int d^4x_L G_n(x_0, x_1, \ldots , x_L).
\end{equation}
In order to compute this expression we use, following Broadhurst\footnote{A modernized version of this derivation, that makes use of the important concept of {\it single-valued multiple polylogarithms}, is contained in \cite{S13}, Example 3.31.} [B], the relation
\begin{eqnarray}
\label{P()}
P_L(x_0, x_{L+1})&:=&\prod_{k=1}^L \int\frac{d^4 x_k}{\pi^2 x_k^2}\prod_{m=0}^L\frac{1}{x_{mm+1}^2} \nonumber \\
&= &\int\frac{d^4x_k}{\pi^2 x_k^2} P_{k-1}(x_0, x_k) P_{L-k}(x_k, x_{L+1})
\end{eqnarray}
which yields
\begin{eqnarray}
\label{PL}
&& x^2 P_L(x, y) = \sum_{n=1}^\infty C_{n, L}(r) r^{n-1} \frac{\sin n\theta}{\sin\theta}, \nonumber \\
&&{\rm for}  \,  r^2 = \frac{y^2}{x^2},  \, \, \frac{x \cdot y}{rx^2} := cos\theta \, , \ \left( \frac{\sin n \theta}{\sin \theta} = C_{n-1}^1 (\cos \theta) \right) , \nonumber \\
&&C_{n, L}(r) = \frac{1}{n^{2L}}\sum_{k=0}^L \binom{2L-k}{L} \frac{(\ell n \, r^{-2n} )^k}{k!} \, .
\end{eqnarray}
Inserting the result in (\ref{resBro}) we find (for $L=n-1, r = 1$):
\begin{equation}
\label{zeta}
{\rm res} \, G_n = 2\pi^{2n}  \binom{2n-2}{n-1} \zeta(2n-3) \, .
\end{equation}  

The notion of residue does not depend on the renormalization procedure or on the above choice of  a seminorm. We shall sketch an alternative calculation which uses instead a norm 
for the analytic regularization of $G_n$,  
\begin{equation}
\label{GR}
G_n^\varepsilon = \left( \frac{R}{\ell} \right)^\varepsilon G_n\ , 
\quad  R = max(r_0, r_1 ..., r_L);
\end{equation}
instead of the integration over ${\mathbb R}^{4L}$ in (\ref{resBro}), here we will only have a sum of integrals over finite simplexes, obtained from the standard one
 \begin{equation}
\label{sec0L}
I: \, r_0 (=1) \geq r_1\geq  ... r_L \ (\geq 0)
\end{equation}
by a permutation of the subscripts $\sigma: (1, 2, ..., L) \rightarrow (\sigma 1, \sigma 2, ..., \sigma_L)$ (the symmetry of the integrand allowing to choose $R=r_0$ and multiply 
the result by $n=L+1$): 
\begin{eqnarray}
\label{eq4.14new}
{\rm res} \, G_n &= &n \int d^4 x_0 \, \delta (r_0 - 1) \int_{r_1 \leq 1} d^4 x_1 \ldots \int_{r_L \leq 1} d^4 x_L \, G_n (x_0 , x_1 , \ldots , x_L) \nonumber \\
&= &(L+1) \sum_{\sigma \in {\mathcal S}_L} {\rm res}_{\sigma} \, G_{L+1} 
\end{eqnarray}
where ${\rm res}_{\sigma} \, G_{L+1}$ is the integral of
\begin{equation}
\label{eq4.15new}
\tilde G_{L+1} (r_1 , \ldots , r_L) := \int_{{\mathbb S}^3} d \omega_0 \ldots \int_{{\mathbb S}^3} d \omega_L \, G_{L+1} (\omega_0 , r_1 \omega_1 , \ldots , r_L \omega_L)
\end{equation}
over the simplex $1 \geq r_{\sigma_1} \geq \ldots > r_{\sigma_L} \geq 0$. It turns out that the angular integral (\ref{eq4.15new}) is a {\it polylogarithmic function} 
whose argument depends on the sector of radial variables (see Appendix). In the case of the standard simplex (\ref{sec0L}) we have:
\begin{eqnarray}
\label{Izeta}
{\rm res}_I \, G_n =  \, n (2\pi^2)^n  \int_0^1 \frac{dr_1}{r_1}... \int_0^{r_{L-1}} \, \frac{dr_L}{r_L} Li_{L-1}(r_L^2)   \, \,  \nonumber \\  
= 2n \pi^{2n} Li_{2L-1}(1) = 2n \pi^{2n} \zeta (2L-1)  \,  = 2n \pi^{2n} \zeta(2n-3). \, \,  \, \, \, 
\end{eqnarray}
Next we observe that there is a subset ${\mathcal S}_L^{(1)}$ of the symmetric group ${\mathcal S}_L$ of $\vert {\mathcal S}_L^{(1)} \vert = 2^{L-1}$ elements $\sigma$ 
yielding the same residue: ${\rm res}_\sigma \, G_n =  \,{\rm res}_I \, G_n$ (given by (\ref{Izeta})). To prove that there are at least $2^{L-1}$ such permutations we 
proceed by induction in $L$ using the relation  
\begin{equation}
\label{n-s}
{\rm res}_\sigma \, G_n =  \,{\rm res}_{n-\sigma} \, G_n \, \,  \mbox{for} \, n-\sigma = (n-\sigma_1, ..., n-\sigma_L), \, \, n=L+1.
\end{equation}
For $L=2$, the first step of the induction, this relation is reduced to the observation that the residue (\ref{Izeta}) for the standard simplex is equal to that for the ``reversed''
one: $r_0\geq r_2\geq r_1$. (This suffices to reproduce the result (\ref{zeta}) for the physically relevant case $n=3: {\rm res} \, G_3 = 12\pi^6 \zeta (3)$.) To go from $n$ to $n+1$
we relable the indices $1, ..., L$ as $2, ..., n(=L+1)$ in each of the $2^{L-1}$ permutations of ${\mathcal S}_L^{(1)}$ and insert $1$ on the left. The resulting $\sigma$'s together with 
the permutations $n+1 - \sigma$ are the required $2^L$ elements of ${\mathcal S}_n^{(1)}$. The rest of the
proof is relegated to the Appendix where the computation of ${\rm res} \, G_n$ for $n=4, 5, 6$ as a sum of integrals over simplexes is also sketched.
     
The above calculations exhibits a general feature that goes beyond primitively divergent graphs: the reduction at each step of the renormalization to an one-dimensional problem 
in a radial scale parameter (while there is no problem in integrating the remaining angular variables). This remark has been used 
systematically in \cite{NST, NST12}; lately it was also exploited in the momentum space picture \cite{BKr}. The appearance of $\zeta$-values in similar computations (that is given a 
motivic interpretation in \cite{BEK}) has been detected in early work of Rosner \cite{R}.  Broadhurst and Kreimer \cite{BrK}, \cite{Kr} related it to the topology of graphs.

\smallskip
\subsection{A possible redefinition of time-ordered products}

As demonstrated in \cite{N09} having carried out (inductively) the renormalization satisfying causal factorization one can, for any local QFT model, reconstruct 
the Epstein-Glaser time-ordered products $T(\phi_1(x_1)... \phi_n(x_n))$ where $\phi_i$ are arbitrary Wick monomials in (derivatives of) the basic fields. In particular, 
{\it the renormalized time-ordered product is translation invariant}:
\begin{equation}
\label{dT}
\partial T(\phi_1(x_1)... \phi_n(x_n)) = T\partial (\phi_1(x_1)... \phi_n(x_n)) \quad
\mbox{for} \quad \partial = \sum_{i=1}^n \frac{\partial}{\partial x_i}\ .
\end{equation}
It is seldom stressed that in practice one uses a stronger requirement: commutation of time-ordering with all derivatives which yields conservation of momentum at each vertex of a Feynman graph (a more restrictive property than the total momentum conservation!). As explained before the resulting renormalization cannot originate from a linear extension map; it satisfies instead the {\it action Ward identity} \cite{DF}. We shall denote the corresponding time-ordered
product by $T_{AWI}$. Happily, given a translation invariant $T$, one can reconstruct $T_{AWI}$ by a {\it finite renormalization}. To this end we introduce a basis of {\it balanced derivatives}, say $\partial_{i i+1}=\frac{\partial}{\partial x_i}-\frac{\partial}{\partial x_{i+1}}, i=1,... n-1$, and set
\begin{equation}
\label{partial}
T_{AWI}(\partial_{12}^{k_1}...\partial_{n-1 n}^{k_{n-1}}\partial^k\phi_1(x_1)...\phi_n(x_n)) =
\partial^k T(\partial_{12}^{k_1}...\partial_{n-1 n}^{k_{n-1}}\phi_1(x_1)...\phi_n(x_n))\ .
\end{equation}
It is easy to verify that for $T$ satisfying (\ref{dT}) $T_{AWI}$ thus defined does commute with all derivatives.

The above remarks illustrate the fact that different consistent approaches to renormalization may have conflicting requirements (like linearity vs. action Ward identity). We see however that to pass from one consistent scheme to another one does not have to redo the infinite renormalization. In the above example we had just to transform one (well defined) time-ordered product into another.

\bigskip

\section{Outlook}
\setcounter{equation}{0}

Quantum field theory which once signaled, according to Freeman Dyson \cite{D72}, a divorce between mathematics and physics, now seems to be the best common playground of the two sciences. Not only did renormalization theory, which was viewed as a liability, become respectable in the above sketched framework. Combined with the idea of dimensional transmutation \cite{F}, it gives the 
best hope for solving the fifth Millennium Prize Problem of the Clay Mathematics Institute (the existence of a mass gap in a pure Yang-Mills theory with a nonabelian, simple, compact gauge group); the inherent calculations of residues/periods is fascinating and attractive to mathematicians - cf. \cite{BrK, B13, BKr, S13} and references cited there. Recent work \cite{St13, AD} indicates that the study of globally conformal invariant QFT models, initiated in \cite{NT}, is unlikely to go beyond composites of free fields. This makes all the more attractive the introduction and study (by a number theorist \cite{B04}) of the shuffle algebra of {\it single valued multiple polylogarithms} (SVMP) which allowed to write down two-loop off shell conformal 4-point functions (in ${\mathcal N}=4$ super Yang-Mills theory) in a closed analytic form \cite{DDEHPS}. Application of conformal invariance to perturbative calculations was enhanced by the discovery of {\it dual conformal invariance} (for momentum-like coordinates) \cite{DHSS, DHKS}. Combined with the standard (x-space) conformal invariance, it yields an infinite dimensional {\it Yangian symmentry}. Recent work of mathematical physicists and algebraic geometers which applies this symmetry and can be traced back from \cite{ABCGPT, GGSVV} suggests that one can construct (up to a common infrared divergent factor that cancels out in the ratios of successive approximations) a conformally invariant scattering matrix of massless "gluons". Similar tools (including SVMP) appear in realistic QCD calculations applicable to current experiments at LHC - see \cite{DDDP}.

It seems to be an excellent time for a new generation of mathematical physicists to enter the scene!

{\bf Acknowledgments}. It is a pleasure to thank Dirk Kreimer for his invitation and hospitality at the Humboldt University, Berlin Conference {\it Quantum Field Theory, Periods and Polylogarithms}, and l'Institut des Hautes \'Etudes Scientifiques, Bures-sur-Yvette for hospitality during the completion of these notes. I thank David Broadhurst, Nikolay Nikolov, Yassen Stanev and Raymond Stora for sharing their insight with me. The author's work has been supported in part by grant DO 02-257 of the Bulgarian National Science Foundation.

%\newpage

%\section*{Appendix. Pascual Jordan (1902-1980)}

\bigskip
 
\section*{Appendix. Sums of integrals over simplices yielding a decomposition of Catalan's numbers}\label{secAA}

According to (\ref{eq4.14new}) (\ref{eq4.15new}) the residue of the wheel graph with $L+1$ spokes is given by
$$
{\rm res} \, G_{L+1} = (L+1) \, \pi^{2(L+1)} \sum_{\sigma \in {\mathcal S}_L} {\mathcal I}_{\sigma} \eqno ({\rm A.1})
$$
where
$$
\label{Is}
{\cal I}_\sigma = \int_0^1 \frac{dr_{\sigma 1}}{r_{\sigma 1}} \int_0^{r_{\sigma 1}} \frac{dr_{\sigma 2}}{r_{\sigma 2}} ... \int_0^{r_{\sigma (L-1)}} \frac{dr_{\sigma L}}{r_{\sigma L}} 
\left(\prod_{i=0}^L \int_{{\mathbb S}^3} \frac{d\omega_i}{\pi^2}\right) G_{L+1} (\omega_0, r_1 \omega_1, ...r_L \omega_L). \eqno ({\rm A.2})
$$
The integral over the spherical angles $\omega_i$ can be computed in $D= 2\lambda+2$ dimensional space-time using the relations
$$
(x_{ij}^2)^{-\lambda} = (r_i^2 + r_j^2 -2 r_i r_j \omega_i \omega_j)^{-\lambda}= \frac{1}{R_{ij}^{2\lambda}} \sum_{n=0}^{\infty} \left( \frac{r_{ij}}{R_{ij}} 
\right)^n C_n^\lambda (\omega_i \, \omega_j) \, ,
$$
$$
\label{eq3.7}
R_{ij} = \max (r_i , r_j) \, , \quad r_{ij} = \min (r_i , r_j) \, , \quad i \ne j \, , \ i,j = 1,2,3.  \eqno ({\rm A.3})
$$
We shall also use the integral formula
$$
\label{eq3.8}
\int_{{\mathbb S}^{2\lambda+1}} d \omega \, C_m^\lambda (\omega_1 \, \omega) \, C_n^\lambda (\omega_2 \, \omega) = \frac{\lambda |{\mathbb S}^{2\lambda +1}|}{n+\lambda} \, 
\delta_{mn} \, C_n^\lambda (\omega_1 \, \omega_2) \, ,  \eqno ({\rm A.4})
$$
where $|{\mathbb S}^{2\lambda+1}| = \frac{2 \,\pi^{\lambda+1}} {\Gamma(\lambda+1)}$ is the volume of the unit hypersphere in $D = 2\lambda + 2$ dimensions.

For the trivial (identity) permutation in $D=4$ dimensions we have, in accord with (\ref{Izeta}), 
$$
{\mathcal I}_1 = 2^{L+1} \int_0^1 \frac{dr_1}{r_1}  \int_0^{r_1} \frac{dr_2}{r_2} \ldots  \int_0^{r_{L-1}} \frac{dr_L}{r_L} \, Li_{L-1} (r_L^2) = 2Li_{2L-1} (1) = 2\zeta (2L-1) 
\eqno ({\rm A.5})
$$
where
$$
Li_k (\xi) = \sum_{n=1}^{\infty} \frac{\xi^n}{n^k} \quad (Li_1 (\xi) = -\ell n (1-\xi) , Li_k (1) = \zeta (k)) \, . \eqno ({\rm A.6})
$$
We constructed in Sec.~4.1 $2^{L-1}$ permutations $\sigma = (\sigma1 , \ldots , \sigma L)$ of $(1,\ldots , L)$ for which ${\mathcal I}_{\sigma} = {\mathcal I}_1$. For $L>2$ there are more permutations and for all $\sigma$'s not covered by that construction the integral (A.1) is a fraction of ${\mathcal I}_1$. If $\sigma_{L-2}$ interchanges $L-2$ with $L-1$ then the corresponding integral ${\mathcal I}_2 := {\mathcal I}_{\sigma_{L-2}}$ is
$$
{\mathcal I}_2 (={\mathcal I}_2(L)) = 2^{L+1} \int_0^1 \frac{dr_1}{r_1} \ldots \int_0^{r_{L-3}} \frac{dr_{L-1}}{r_{L-1}}  \int_0^{r_{L-1}} \frac{dr_{L-2}}{r_{L-2}}  \int_0^{r_{L-2}} \frac{dr_L}{r_{L}} \, 
$$
$$
Li_{L-1} \left( \frac{r_{L-2}^2 \, r_L^2}{r_{L-1}^2} \right) = \zeta(2L-1)
= \frac12 \ {\mathcal I}_1 \, . \eqno ({\rm A.7})
$$
There are $(L-2) \, 2^{L-2}$ permutations $\sigma$ which yield the same value (A.7) of the integral ${\mathcal I}_{\sigma}$. The remaining $\sigma \in {\mathcal S}_L$ 
(that appear for $L>3$) give smaller fractions of ${\mathcal I}_1$. For even $L=2\ell$ the smallest such fraction is\footnote{This and some subsequent results are obtained by Javor Boradjiev (in preparation).} $\ell!^{-2}$; for $L= 2\ell +1$ it is $(\ell! (\ell+1)!)^{-1}$. The corresponding multiplicities are given by $2\ell!^2$ and $\ell! (\ell+1)!$, respectively. A representative permutation with minimal value of the integral ${\mathcal I}_{\sigma}$ in each case is given by $\sigma=(2, 4, ... 2\ell, 1, 3, ..., 2\ell \mp 1)$.  Here are three examples that exhaust (along with (A.5), (A.7) and (A.8)) the possible values of ${\mathcal I}_{\sigma}$ for $L \leq 5$:
$$
{\mathcal I}_{(2314)} = 2^{5} \int_0^1 \frac{dr_2}{r_2} \int_0^{r_2} \frac{dr_3}{r_3}  \int_0^{r_3} \frac{dr_1}{r_1}  \int_0^{r_1} \frac{dr_4}{r_{4}} \, Li_3 \left( \frac{r_1^2 \, 
r_4^2}{r_2^2} \right) 
$$
$$
= \frac12 \ \zeta (7)  = \frac14 \, {\mathcal I}_1  \eqno ({\rm A.8})
$$
(there are $2^{L-4} (L-3)(L+4)$ different permutations $\sigma \in {\mathcal S}_L, L>3$ yielding the same ${\mathcal I}_{\sigma}=\frac14 {\mathcal I}_1$; in particular, the eight permutations for $L=4$ are obtained from the pair $(2314) (2413)$ by applying $(\sigma_{L-1} =) 
\sigma_3$ : 

\noindent $(a, b, c, d) \to (a, b, d, c)$ and $n-\sigma$ (4.17));
$$
{\mathcal I}_{(14235)} = \frac14 \ \zeta (9) = \frac18 \, {\mathcal I}_1\, , \eqno ({\rm A.9})
$$
$$
{\mathcal I}_{(42135)} = \frac16 \ \zeta (9) = \frac{1}{12} \, {\mathcal I}_1\, . \eqno ({\rm A.10})
$$
There are $2^{L-4} (L-4)[\frac{L(L+13)}{6}+1]$ permutations $\sigma \in {\mathcal S}_L, L>4 $ giving the value $\frac18 {\mathcal I}_1$ for $I_{\sigma}$, and $2^{L-3} 3(L-4)$ other which give $\frac{1}{12} \, {\mathcal I}_1$.

\smallskip

In view of (\ref{zeta}) we have in general
$$
\sum_{\sigma \in {\mathcal S}_L} {\mathcal I}_{\sigma} = 2 C_L \, \zeta (2L-1) \, , \quad C_L = \frac{1}{L+1} \left({2L \atop L} \right) \eqno ({\rm A.11})
$$
where $C_L$ are the {\it Catalan numbers}. Let $s$ stand for the set of permutations $\sigma$ with the same value of the integral $I_{\sigma}$:

$$
{\mathcal I}_\sigma = \frac{1}{n_s} \, {\mathcal I}_1 \, (\sigma \in s), \, 
n_s = 1, \, 2, \, 4, \, 8, \, 12, \, 16, \, 24, ...  .   \eqno ({\rm A.12})
$$
%(2^4 + 3 \times 2^3 + 36 + 32 + 12 = 5 \, !) \, . 

One verifies that the number $n_s$ divides the cardinality $\vert  s \vert$ of the subset $s(=s^{(L)})$. Defining 

$$
N_L (n_s)=\vert  s \vert/n_s, \, \, \mbox{so that} \, \, C_L = \sum_s N_L(n_s) \eqno ({\rm A.13})
$$
we find (setting $(x)_+=\frac12 (x+|x|)$)
$$
N_L (1) = 2^{L-1}, \, N_L(2) = 2^{L-3}(L-2)_+, \, N_L(4)=2^{L-6}(L-3)_+(L+4),  \, 
$$
$N_L(8)=2^{L-7}(L-4)_+ [\frac{L(L+13)}{6}+1], N_L(12) = 2^{L-5}(L-4)_+, ...,$ \, 
$$
N_L(24) = 2^{L-7} (L-5)_+ (L+2), ... , N_L(36) = 2^{L-5} (L-5)_+. \, \eqno ({\rm A.14})
$$  

There is an intriguing alternative way to define the positive integers $N_L (n_s)$ (A.13). 
In order to state it we first recall one of the many characteristics of the Catalan numbers. $C_L$ is equal to the number of permutations ${\sigma} \in {\mathcal S}_L$ which avoid configurations of the type
$$
\ldots b \ldots c \ldots a \ldots \quad \mbox{for} \quad a,b,c \in {\mathbb N} \, , \quad 1 \leq a < b < c \leq L \, . \eqno ({\rm A.15})
$$
If we denote by $a_s \subset s$ the subset of allowed permutations in $s$ (that avoid the configurations (A.15)) then it turns out that $N_L(n_s)$ is nothing but the number of elements 
$|a_s|$ in $a_s$. 

\smallskip
 
The above formulae allow to recover Eqs. (A.11) (A.13) for $L\leq 5$. One can also verify it numerically for higher $L$'s. It may be instructive to find a general formula for $N_L(n_s)$, thus obtaining a decomposition of the Catalan number $C_L$ for any $L$ into a sum of positive integers characterizing the different contributions of the integrals ${\mathcal I}_\sigma$. 
%{\bf References}

\end{document}